# Superconductivity in Undoped BaFe$_2$As$_2$ by Tetrahedral Geometry Design

J. H. Kang[1], J.-W. Kim[2], P. J. Ryan[2], L. Xie[3], L. Guo[1], C. Sundahl[1], J. Schad[1], N. Campbell[4], Y. G. Collantes[5], E. E. Hellstrom[5], M. S. Rzchowski[4], and C. B. Eom[1]*

[1]Department of Materials Science and Engineering, University of Wisconsin-Madison, Madison, WI, USA
[2]Advanced Photon Source, Argonne National Laboratory, Argonne, IL, USA
[3]Southern University of Science and Technology, Shenzhen, China
[4]Department of Physics, University of Wisconsin-Madison, Madison, WI, USA
[5]Applied Superconductivity Center, National High Magnetic Field Laboratory, Florida State University, 2031 East Paul Dirac Drive, Tallahassee, FL 32310, USA
School of Physical Sciences, Dublin City University, Dublin 9, Ireland

*e-mail: eom@engr.wisc.edu

**Fe-based superconductors exhibit a diverse interplay between charge, orbital, and magnetic ordering[1-4]. Variations in atomic geometry affect electron hopping between Fe atoms[5,6] and the Fermi surface topology, influencing magnetic frustration and the pairing mechanism through changes of orbital overlap and occupancies[7-11]. Here, we experimentally demonstrate a systematic approach to realize superconductivity without chemical doping in BaFe$_2$As$_2$, employing geometric design within an epitaxial heterostructure. We control both tetragonality and orthorhombicity in BaFe$_2$As$_2$ through superlattice engineering, which we experimentally find to induce superconductivity when the As-Fe-As bond angle approaches that in a regular tetrahedron. This approach of superlattice design could lead to insights into low dimensional superconductivity in Fe-based superconductors.**

Epitaxial thin films and heterostructures can fine tune the atomic arrangement of complex materials and subsequently enhance the superconducting transition temperature($T_c$), as recently demonstrated in single-layer FeSe[12]. More complex high-$T_c$ superconductors such as Co-doped $BaFe_2As_2$ have technical challenges such as structural and chemical disorder at surfaces and interfaces, but also offer the promise of inducing and enhancing superconductivity with structural control. Highly ordered and controlled interfaces can host superconducting properties superior to that of bulk materials[13-21]. Here, we present the atomic engineering to control and measure interfacial driven superconductivity in Fe-based compounds. The atomically sharp interfaces in undoped $BaFe_2As_2$ (Ba-122)/$SrTiO_3$ (STO) superlattice systems offer advancement in the precise control of atomic structures.

The relation between atomic structure and electronic properties of superconducting $Ba(Fe_{1-x}Co_x)_2As_2$ and the parent compound $BaFe_2As_2$ has emerged as the key to understanding the complex nature of these materials, including the low-temperature superconducting state. The nematic phase arising at the structural transition temperature ($T_s$) breaks the 90° rotation symmetry of the high-temperature phase, necessarily exhibiting structural, orbital, and spin-driven nematic order[4,7], with antiferromagnetic order setting in at a temperature $T_n < T_s$. Increased Co doping pushes the magnetic transition further below the structural orthorhombic (nematic) transition, and suppresses both associated order parameters. This correlates with the emergence of the low-temperature superconducting phase, which competes with static magnetic order. The highest superconducting transition temperature $T_c$ is obtained at the Co doping for which not only is the orthorhombicity smallest, but also where the Fe-As bonds form a symmetric tetrahedral arrangement.

An important question is the driving force behind these effects, whether spin, charge or structural effects. Substitutional doping affects electronic interactions, which in turn influence other orders in the material. Here we demonstrate the separate importance of orthorhombicity and tetrahedral coordination by inducing superconductivity in the undoped parent compound $BaFe_2As_2$ by strain and symmetry control. We accomplish this by alternating ultrathin layers of $BaFe_2As_2$ with $SrTiO_3$ in an epitaxial superlattice grown on a $CaF_2$ substrate. The substrate provides initial biaxial strain, and the $SrTiO_3$ layers maintain that strain throughout the superlattice. The square template of both of these layers suppresses the orthorhombicity of low-temperature $BaFe_2As_2$, with the thinnest $BaFe_2As_2$ layers remaining almost fully tetragonal to low temperature. The undoped superlattice becomes superconducting, and the transition temperature increases with decreasing $BaFe_2As_2$ orthorhombicity, and with decreasing differences between Fe-As bond angles. We argue that structural control is an important contributor to superconductivity in Fe based superconductors.

In parent Ba-122 bulk single crystals (Fig. 1a), the Fe sublattice undergoes a symmetry-breaking magnetostructural phase transition at low temperature, distorting the $FeAs_4$ tetrahedron and altering all bond angles $\alpha$, $\beta$, and $\gamma$, and changing the overall symmetry from tetragonal to orthorhombic. The magnetic ordering in the quasi-2D plane of the distorted $FeAs_4$ tetrahedra is believed to suppress superconductivity in bulk Ba-122[22,23]. However, we demonstrate superconductivity by tuning the system towards a higher-symmetry structure, namely a regular tetrahedral arrangement[24] in high-quality epitaxial Ba-122/STO superlattice films on (001) $CaF_2$ substrates (see Methods and Supplementary Sections 1.1 and 1.2). This is achieved via the manipulation of in-plane symmetry (orthorhombicity at low temperature) and the ratio between out-of-plane and in-plane lattice parameter (tetragonality at high temperature) of the Fe sublattice structures (Figure 1b).

The superlattice system provides a mechanism to control the $FeAs_4$ tetrahedron of the Ba-122 layers for emergent superconductivity. First, bulk Ba-122 has anisotropic in-plane lattice constants associated with the tetragonal-to-orthorhombic transition, causing the $\beta$ and $\gamma$ bond angles to differ. We found that the orthorhombic transition is suppressed by in-plane clamping arising from epitaxy with STO, leading to equal $\beta$ and $\gamma$. Secondly, tetragonal Ba-122 without the orthorhombic distortion still has a bond angle difference between $\alpha$ and $\beta$, as they are set by the in-plane lattice and the anion (Arsenic) $z$ height[9,25] (see Methods and Supplementary Section 2). Since in bulk Ba-122 $\alpha$ is larger than $\beta$ as shown in Figure 1a, reducing the in-plane lattice and expanding the anion $z$ height drives the bond angle $\alpha$ towards $\beta$, through bi-axial compressive strain imposed by $CaF_2$ substrates. The regular tetrahedron can be achieved by tetragonal elongation where all bond angles approach 109.5° (Figure 1b).

The reduction of Ba-122 thickness ($t$) in the superlattice structure decreases orthorhombic distortion and increases $c$-elongated tetragonality because the rigidity from STO transmits the compressive strain and the square symmetry more effectively through throughout the Ba-122 film (Fig. 1c). However the clamping effect on both the top and bottom of the Ba-122 layer is insufficient to fully suppress the orthorhombic transition. The superlattice Ba-122 layers experience a strain gradient (showing a broad superconducting transition), but the average values of structural and electrical properties tuned by the Ba-122 dimensionality clearly demonstrate the influence of tetrahedral geometry on $T_c$.

We performed atomic-resolution scanning transmission electron microscopy (STEM) combined with energy-dispersive X-ray spectroscopy (EDX) to investigate the microstructure, chemistry and interfacial structure of the superlattice system (Fig. 2). Figure 2a shows a cross-sectional high-angle annular dark field (HAADF) image of the 12-layer Ba-122/STO superlattice structure with a total thickness of 260 nm (See Methods). Bright and dark layers in the image correspond to 7 nm thick Ba-122 and 14 nm thick STO layers, respectively. The thickness of each layer and the modulation wavelength ($\Lambda$) are in accordance with the structural information determined by satellite peaks in synchrotron X-ray diffraction (XRD) patterns (See supplementary Section 3). Figure 2b shows a high-resolution HAADF image of the interfacial region of the STO/Ba-122/STO structure (black dotted box in Fig. 2a). Individual atoms of the Ba-122, STO, and interfaces are distinguishable (inset of Fig. 2b). The Ba-122/STO heterostructure has atomically sharp interfaces at both the top and bottom regions.

Interfacial imperfections such as chemical intermixing between Ba-122 and STO are investigated with energy dispersive X-ray spectroscopy (EDX). The elemental mapping by EDX shows there is no severe chemical reaction, diffusion or intermixing between Ba-122 and STO layers, and we can clearly see the layered distribution of Sr and Fe/As atoms in STO and Ba-122, respectively (Fig. 2c). The EDX line scan indicates that the bottom interface of the Ba-122 layer has less interdiffusion/intermixing than the top (Fig. 2d). Previous theoretical calculations on the STO/Ba-122 interface support suppressed intermixing at the bottom interface, as the Ba layer is thermodynamically stable on $TiO_2$-terminated STO with an atomically sharp interface[26]. In contrast, there is a small amount of As and Fe atoms missing in Ba-122 underneath the top interface (red and blue arrows) due to Ti diffusion (purple arrows). Ba diffusion into the STO is also found above the interface (black arrows). The EDX scan indicates that a Ba/Sr interfacial layer on the As layer[27,28] is unstable during the STO growth, allowing a small degree of intermixing at the interface. Since the intensity of Ti in the Ba-122 layer is from Ba-$L_\alpha$ edges which overlap with Ti-$K_\alpha$ in EDX (indicated by an asterisk in Figure 2d), we confirm that there are no Ti and O atoms in the Ba-122 layer with electron energy loss spectroscopy (See Supplementary Section 4).

The detailed structure of superlattice Ba-122 was investigated by synchrotron X-ray diffraction in Fig. 3 (See Methods and Supplementary Section 5.1, 5.2 and 5.3). We measured the Ba-122 (2 2 8) reflections to examine the tetragonal-to-orthorhombic transition[29] as a function of temperature. As temperature decreases, the peak broadens due to the orthorhombic distortion below the phase transition temperature (See Supplementary Section 5.2 and 5.3). The reflection was fitted with two peaks using the full-width half maximum (FWHM) of the tetragonal phase taken at room temperature represented by the purple and dark blue curves in Fig. 3a. The orthorhombic distortion is clearly shown in the thicker Ba-122, whereas it is reduced in the 3.5 nm thick Ba-122 layers. From the fitted peak positions of (2 2 8) and (0 0 L) reflections, we calculate orthorhombic $a$, $b$ and $c$ lattice parameters. The temperature-dependent in-plane lattice parameters are shown for different Ba-122 thicknesses in Fig. 3b. The $SrTiO_3$ layer thickness is kept constant at 14nm, and all superlattices consist of 12 bilayers. Higher compressive bi-axial strain and the larger c-axis elongation are observed in the thinner Ba-122. As the Ba-122 is grown thinner, the structural transition temperature ($T_s$) shifts lower (indicated by arrows), and the difference between in-plane lattice parameters becomes smaller.

We performed anomalous X-ray scattering at the As-$K$ edge to extract the precise local structure of the $FeAs_4$ tetrahedron. The extent of intensity variation in energy scans of the (0 0 L) at the absorption edges reflects the amount of As scattering contribution to the diffraction intensity, which is directly coupled to the $z$ position (See Methods and Supplementary Section 5.4). Combining the lattice parameters with the As relative $z$ position, the complete $FeAs_4$ configuration is obtained. Figure 3c demonstrates that the As-Fe-As bond angles are controlled by the Ba-122 thickness. The three bond angles $\alpha$, $\beta$, and $\gamma$ in the tetrahedron are all different in thicker Ba-122 due to the orthorhombic symmetry. As the Ba-122 thickness decreases, the difference between the bond angles reduces until finally there is a negligible difference at 3.5 nm showing an ideal tetrahedron bond angle[10,24] of 109.5°. The most symmetric tetrahedron is obtained via both the suppression of orthorhombicity by the clamping effect and the manipulation of the tetragonality by the compressive bi-axial strain. Therefore, the superlattice design with the Ba-122 thickness control provides an explicit variable to realize the regular tetrahedron in parent Ba-122 (See Supplementary Section 5.5).

To investigate the influence of tetrahedral geometry on superconducting $T_c$, temperature-dependent resistivity (Fig 4a) was measured in a van der Pauw geometry. The data show resistive anomalies corresponding to the structural and magnetic transitions. We assigned the structural phase transition temperature ($T_s$) to the onset of the first derivative of the resistivity, and the magnetic phase transition temperature ($T_n$) to the maximum in the first derivative of the resistivity with respect to temperature[23,29,30] (See Supplementary Section 6.1 and 6.2). Reduction of $T_s$ with decreasing the Ba-122 thickness is consistent with the suppression of the orthorhombic distortion characterized by X-ray measurements as shown in Figure 3b. Interestingly, $T_n$ also moves to lower temperatures as the structural distortion is suppressed, in agreement with previous reports of antiferromagnetic ordering suppression in parent Ba-122[31,32]. By suppressing the antiferromagnetically ordered orthorhombic phase, we observed emergent superconductivity and increased $T_c$ up to 9 K in the superlattice sample with thinner Ba-122 layers (inset of Figure 4a). It is clearly shown that the $T_c$ increases from 0 K to 9 K as the Ba-122 thickness decreases from 20 nm to 7 nm. The 3.5 nm layer thickness superlattice did not reach zero resistance above 2K, likely due to localization of Cooper pairs by scaling dimensions or scattering from the interface[33-35] (See Supplementary Section 7.1).

Figure 4b shows the temperature dependent resistivity of Ba-122$_{7\,nm}$/STO$_{14\,nm}$ superlattice in various magnetic field strengths perpendicular to the film surface. It shows a resistive broadening and a lowering of the superconducting transition with increasing magnetic field. However, the current-voltage characteristics in all the superlattice samples were nonlinear, even under high magnetic fields (See Supplementary Section 7.2), indicating that vortices are strongly pinned in the Ba-122 layer (inset of Figure 4b), different from weak pinning in other low-dimensional superconductors[36]. A zero-field-cooled magnetization $T_c$ was measured to show clear diamagnetic signal by superconducting quantum interference device (SQUID) magnetometer measurements (See Supplementary Section 7.3).

Figure 5 shows the temperature-thickness phase diagram of the parent Ba-122 in the superlattice system obtained from our experimental data. Structural ($T_s$) and magnetic ($T_n$) phase transition temperatures and superconducting critical temperature ($T_c$) are shown. Note that $T_s$ was extracted from the bifurcation point of the in-plane lattice constants measured by synchrotron XRD, and $T_n$ was extracted from the maximum in the first derivative of the resistivity with respect to temperature (See Supplementary Section 6.1 and 6.2). The structural transition does not exhibit a 1$^{st}$ order phase transition that is found in Ba-122 single crystals[23,29,30] and $T_n$ deviates considerably from $T_s$, similar to the effects of electron doping. As the Ba-122 thickness decreases to lower dimensions, both $T_s$ and $T_n$ decrease, indicating a suppression of both nematic and magnetic ordering due to bi-axial strain. While $T_s$ and $T_n$ decrease, $T_c$ increases because the broken-symmetry phase, which maintains magnetic ordering and competes with superconductivity, diminishes.

The atomic engineering demonstrated here enables precise control of the $FeAs_4$ tetrahedral geometry, and provides a platform to understand the connection between local structure and superconductivity. Reduction of the Ba-122 layer thickness manipulates the tetrahedra via systematic control of tetragonal and orthorhombic structures, and results in an enhanced superconducting $T_c$. We believe that the superlattice approach provides a pathway to overcome the limited ability to $T_c$ control in the ultrathin films of the Fe-based superconductors[16,37]. Furthermore, this study will open up new opportunities to study low-dimensional superconductivity and to explore novel phenomena in Fe-based superconductors.

**Methods**

Methods, including statements of data availability and any associated accession codes and references, are available.

**Correspondence Information**

Correspondence and requests for materials should be addressed to C.B.E. (eom@engr.wisc.edu).

**Acknowledgements**

This work was supported by the US Department of Energy (DOE), Office of Science, Office of Basic Energy Sciences (BES), under award number DE-FG02-06ER46327. This research used resources of the Advanced Photon Source, a U.S. Department of Energy (DOE) Office of Science User Facility operated for the DOE Office of Science by Argonne National Laboratory under Contract No. DE-AC02-06CH11357. A portion of the work was done at the NHMFL, which is supported by NSF-DMR-1644779 and by the State of Florida. We thank P. J. Hirschfeld, R. Valenti, I. I. Mazin, V. Borisov and Ian Fisher for helpful discussion.

**Author contributions**

J.H.K. and C.B.E. conceived the project. J.H.K. synthesized the samples and carried out the X-ray diffraction and transport measurements. J.W.K. and P.J.R. performed synchrotron X-ray diffraction and resonant scattering. L.G., C.S., and J.S. helped with the synchrotron experiments. L.X. carried out STEM/EDX analysis. N.C. and M.S.R. measured electromagnetic and magnetic properties. Y.G.C and E.E.H synthesized the target. J.H.K., C.B.E., M.S.R., J.W.K., and P.J.R prepared the manuscript. All the authors discussed the results and commented on the manuscript. C.B.E. directed the overall research.

**Competing interests**

The authors declare no competing interests.

**Additional information**

Supplementary information is available for this paper. Reprints and permissions information is available at www.nature.com/reprints. Correspondence and requests for materials should be addressed to C.B.E. Publisher's note: Springer Nature remains neutral with regard to jurisdictional claims in published maps and institutional affiliations.

## METHODS

**Epitaxial growth.** Epitaxial Ba-122/STO superlattices (see Supplementary Sections 1.1, 1.2 and 3) were grown on (001) $CaF_2$ single-crystal substrates by pulsed laser deposition with a KrF (248 nm) ultraviolet excimer laser at 740ºC. The base pressure before the deposition was $3\times10^{-5}$ Pa, and the deposition took place at $3\times10^{-4}$ Pa because of the degassing of the heater. Temperature-dependent in-plane lattice change of the Ba-122 is driven by the thermal expansion coefficient of $CaF_2$ substrate (see Supplementary Section 5.1). We found that fluoride substrates with higher thermal expansion coefficient (CTE) are more effective to provide compressive bi-axial strain rather than oxide substrates. Note that the CTE of $CaF_2$ is $1.9\times10^{-5}$/K at 300K, higher than that of Ba-122 with $1.0\times10^{-6}$/K.

**STEM HAADF & EDX experiments.** Atomic-resolution STEM high-angle annular dark field (HAADF) experiments were carried out on a double aberration-corrector transmission electron microscope, which was operated at 300 kV and has a resolution of ~0.6 Å. A series of STEM HAADF images were obtained at a resolution of 1024×1024 pixels with the dwell time for each pixel ~0.5 µs. Then the acquired images were summed by removing the drift between each image using the cross-correlation method. EDX spectra were acquired by Super-X detectors and processed in software Velox.

**Synchrotron X-ray techniques.** High-resolution X-ray diffraction and resonant X-ray scattering were carried out at beamline 6-ID-B of the Advanced Photon Source (APS) at Argonne National Laboratory. Although the film might have a small strain gradient along the vertical direction, the X-ray diffraction measures an average value of lattice parameters. Orthorhombicity is examined by the reciprocal space scan across the 228 peak. The peak splitting associated with the orthorhombic transition was determined by two-peaks fitting with a fixed reference FWHM obtained from the peak width in the tetragonal phase at room temperature in order to take into account the natural broadening effect due to the Ba-122 layer thickness variant. Anomalous X-rays scattering as As-*K* edge was performed to measure the relative As position in the unit cell. In symmetry group I4/mmm, only As atom is at the Wyckoff position 4e which takes unspecified *z* position in the unit cell. For the bulk crystal case under hydrostatic pressure, only As relative *z* position varies in published diffraction data[1-3]. Since the As contribution on diffraction intensity of (0 0 L) Bragg reflection changes by relative *z* position, deriving As portion of diffraction intensity using anomalous X-ray scattering can simply determine the As position.

## Data Availability

The data that support the findings of this study are available from the corresponding author upon reasonable request.

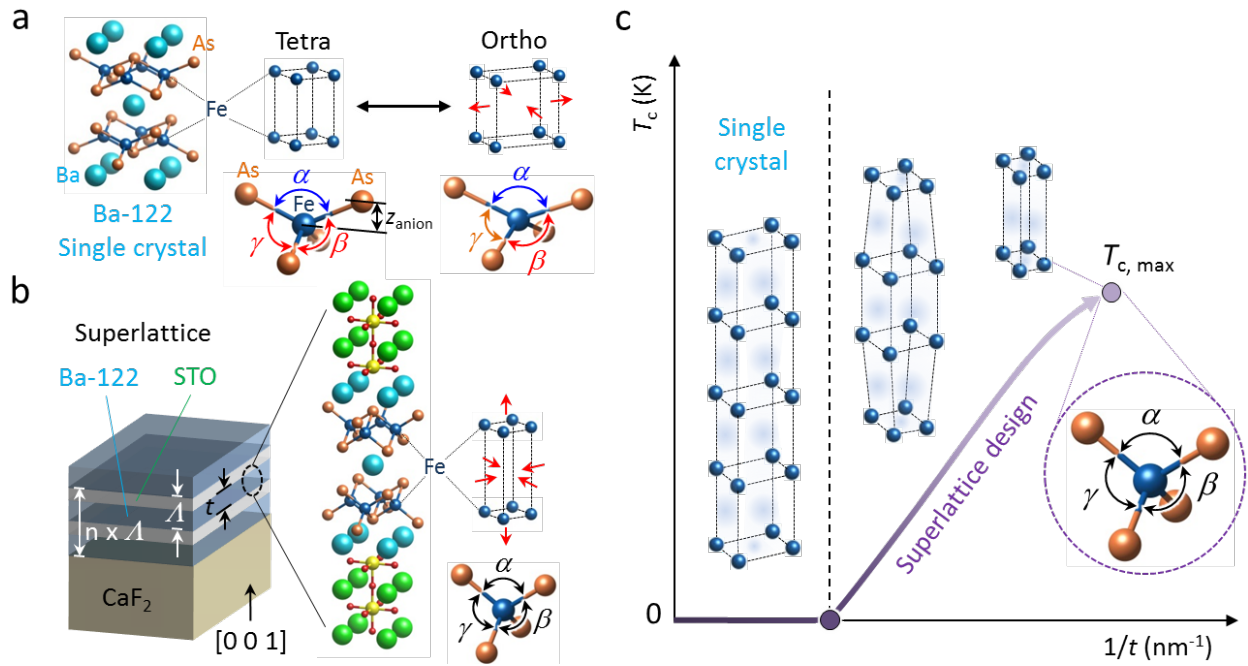

**Figure 1 | Emergent superconductivity in parent Ba-122 by superlattice design. a,** Fe sublattice and tetrahedral geometry in Ba-122 bulk single crystal below magneto-structural phase transition temperature. **b,** Superlattice design for the control of Fe sublattice structure and $FeAs_4$ tetrahedron in the presence of tetragonal-to-orthorhombic transition. **c,** Emergent superconductivity by proximity to the regular tetrahedron via reduction of the Ba-122 layer thickness $t$ in a Ba-122 / $SrTiO_3$ superlattice.

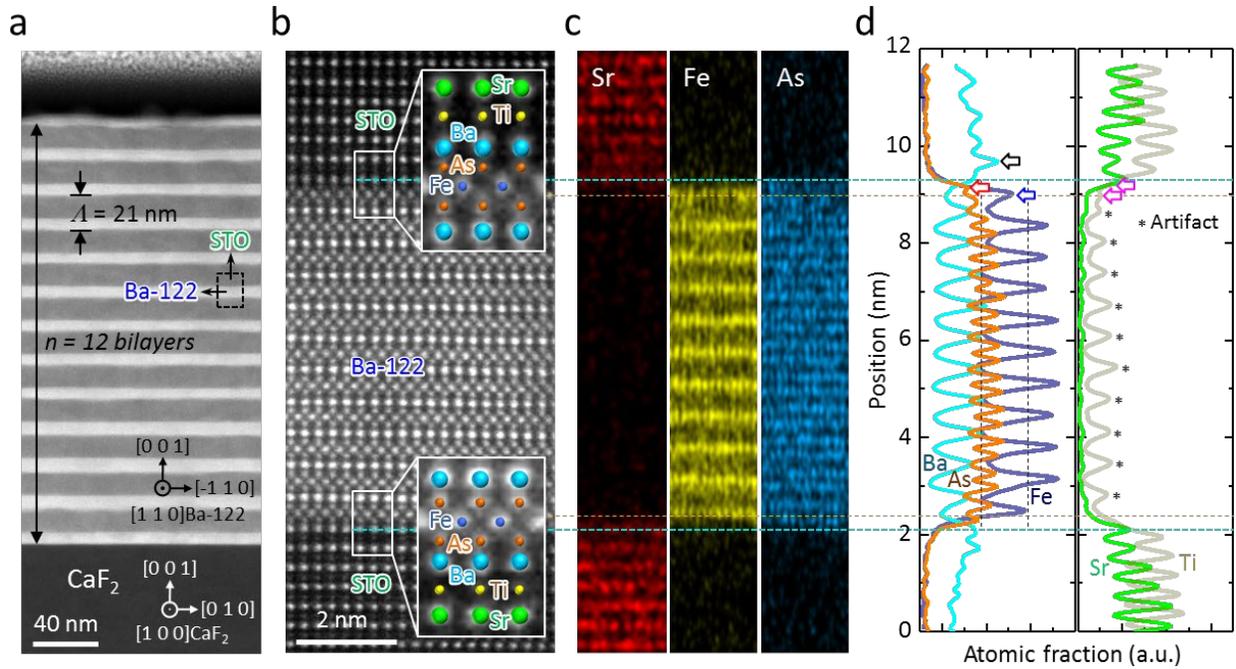

**Figure 2 | Microstructure and atomic arrangement in Ba-122/STO superlattice thin films. a,** HAADF image of 12 layers of Ba-122$_{7\,nm}$/STO$_{14\,nm}$ superlattice on CaF$_2$ substrate. **b,** High-resolution HAADF image of <1 1 0> projection of the Ba-122/STO heterointerfaces. **c,** EDX line scan across an STO insertion layer, a Ba-122 layer, and the interface of a Ba-122/STO. **d,** Atomic fraction extracted from the EDX line scan showing the layered distribution of each element.

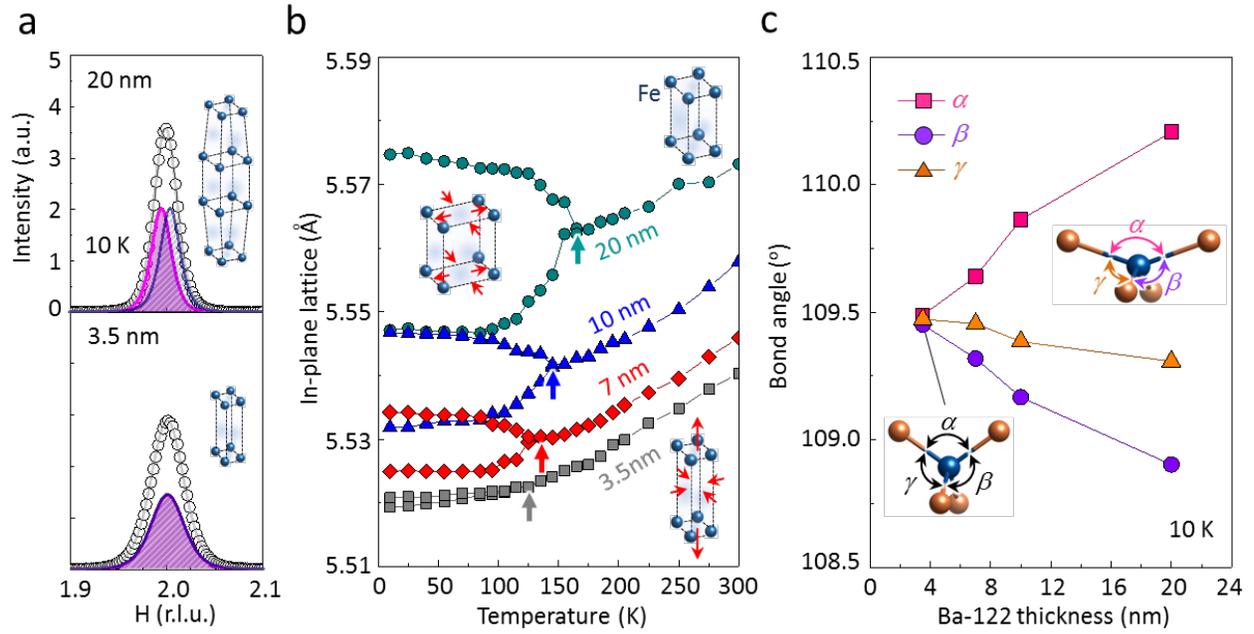

**Figure 3 | Atomic structure of Ba-122 in the superlattice investigated by synchrotron XRD. a,** Tetragonal-to-orthorhombic structural change in (2 2 8) reflection by X-ray diffraction. **b,** In-plane lattice parameters as a function of temperature. **c,** FeAs$_4$ bond angles tuned by Ba-122 thickness.

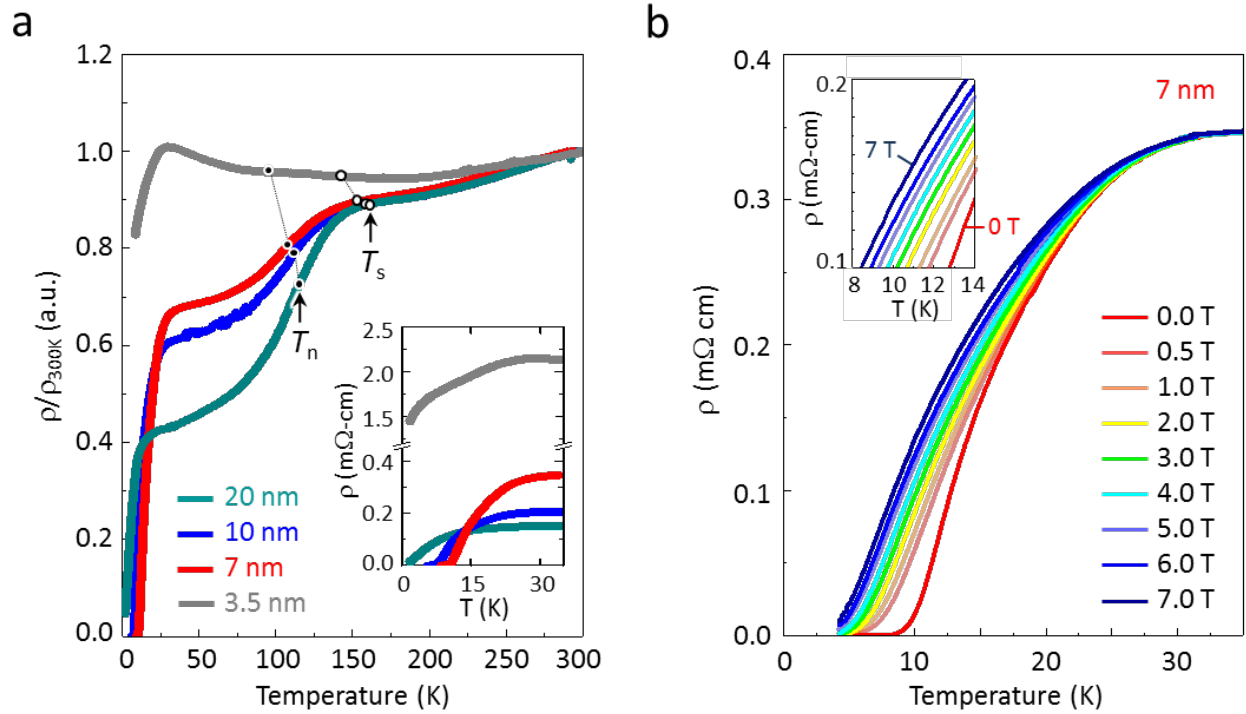

**Figure 4 | Superconducting properties of Ba-122/STO superlattice. a,** Temperature-dependent resistivity with different Ba-122 thickness. $T_s$ and $T_n$ are structural phase transition and the magnetic phase transition temperature, respectively. **b,** Superconducting resistive transition with different magnetic field perpendicular to sample surface.

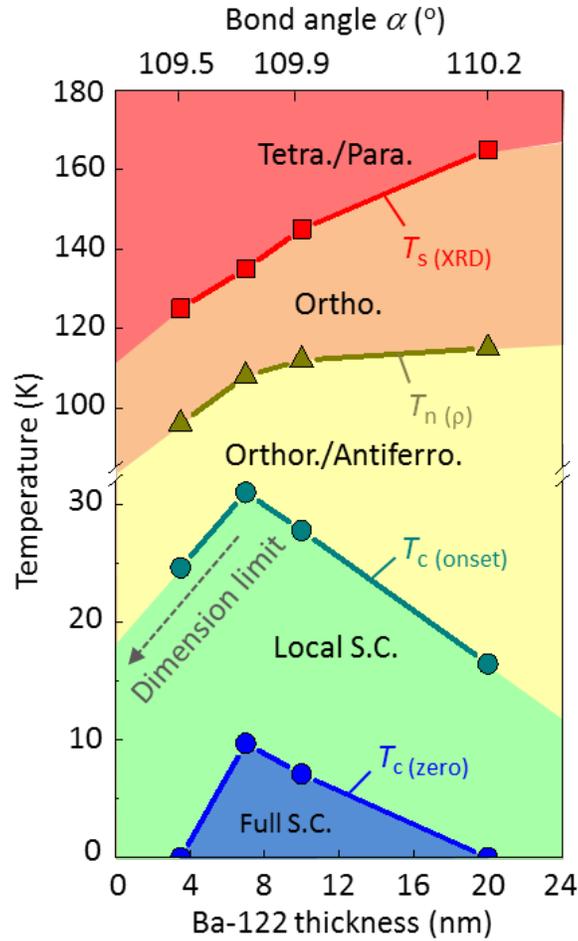

**Figure 5 | Temperature versus dimensionality phase diagram of Ba-122 in the superlattice system.** Broken-symmetry phases are suppressed by dimension control of Ba-122 and superconductivity is enhanced at reduced dimensions. However, superconductivity is weakened due to the dimension limit.

# Supplementary Information

## Superconductivity in Undoped BaFe$_2$As$_2$ by Tetrahedral Geometry Design

J. H. Kang[1], J.-W. Kim[2], P. J. Ryan[2], L. Xie[3], L. Guo[1], C. Sundahl[1], J. Schad[1], N. Campbell[4], Y. G. Collantes[5], E. E. Hellstrom[5], M. S. Rzchowski[4], and C. B. Eom[1]*

[1]Department of Materials Science and Engineering, University of Wisconsin-Madison, Madison, WI, USA
[2]Advanced Photon Source, Argonne National Laboratory, Argonne, IL, USA
[3]Southern University of Science and Technology, Shenzhen, China
[4]Department of Physics, University of Wisconsin-Madison, Madison, WI, USA
[5]Applied Superconductivity Center, National High Magnetic Field Laboratory, Florida State University, 2031 East Paul Dirac Drive, Tallahassee, FL 32310, USA

*e-mail: eom@engr.wisc.edu

# 1. Epitaxial arrangement of Ba-122/STO superlattices

## 1.1. Out-of-plane epitaxial arrangement

Out-of-plane $\theta$-$2\theta$ XRD patterns show only (0 0 L) reflections of the Ba-122 and STO layers in the superlattices (Figure S1). Rocking curves of the (0 0 4) reflection were measured to determine the out-of-plane mosaic spread and crystalline quality. The full width half maximum (FWHM) of the rocking curve is in the range of 1.1º to 1.2º. Even 3.5 nm thick Ba-122 still exhibits good epitaxy shown in a synchrotron XRD scan (Figure S4).

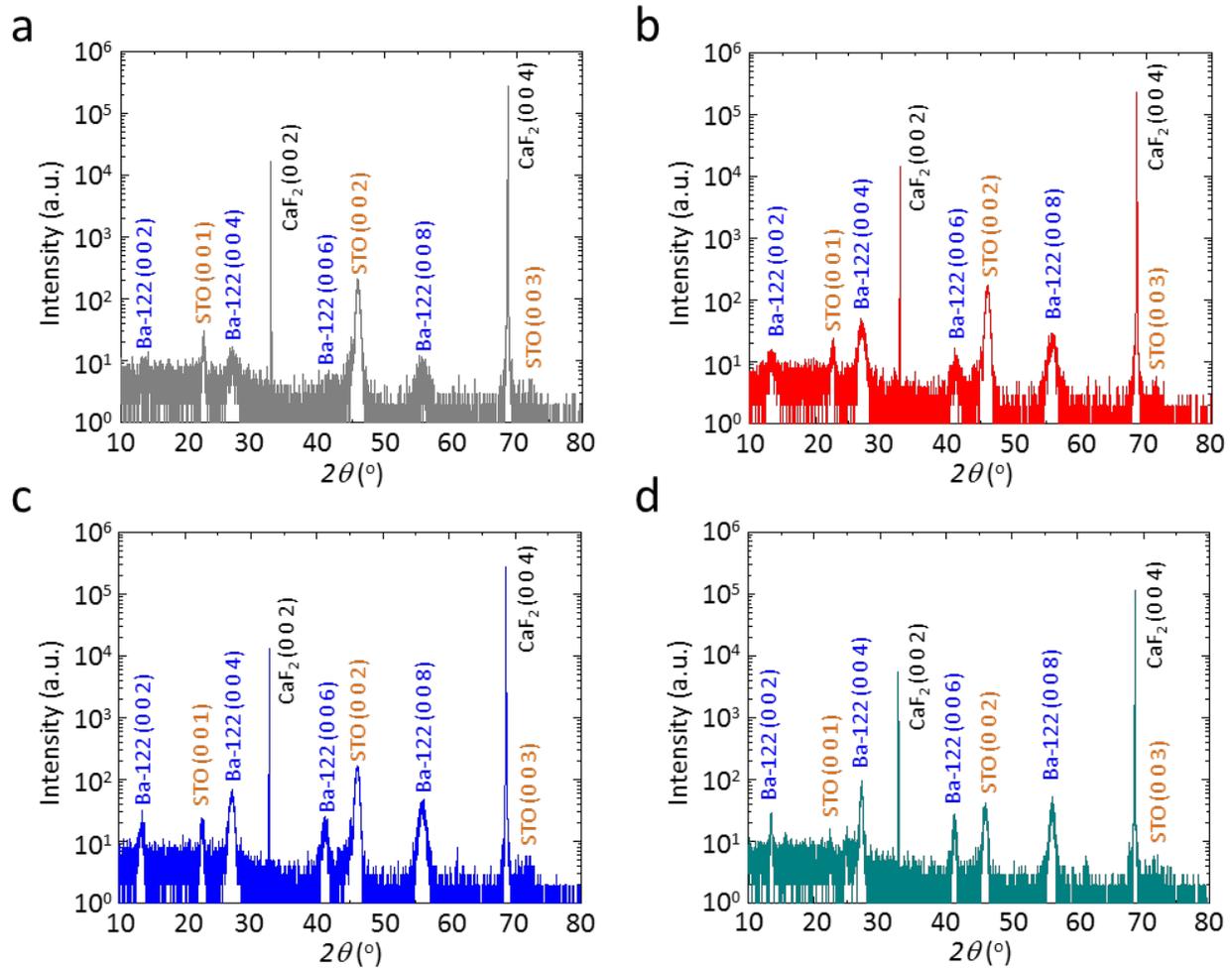

**Fig. S1. Epitaxial quality of Ba-122/STO$_{14\ nm}$ superlattice.** Out-of-plane $\theta$-$2\theta$ XRD patterns of the superlattice on (001) CaF$_2$ substrate with 14 nm thick STO and the Ba-122 with the thicknesses of a) 3.5 nm b) 7 nm, c) 10 nm and d) 20 nm.

## 1.2. In-plane epitaxial arrangement

Azimuthal ϕ scans exhibits in-plane epitaxial quality of the superlattices that there are sharp and strong peaks every 90º characteristic of a truly epitaxial film with excellent in-plane texture. Note that there is an in-plane 45º rotation of the Ba-122 and STO layers with respect to the $CaF_2$ substrate indicating that the [1 0 0] direction of the Ba-122 and STO is parallel to the [1 1 0] direction of the $CaF_2$.

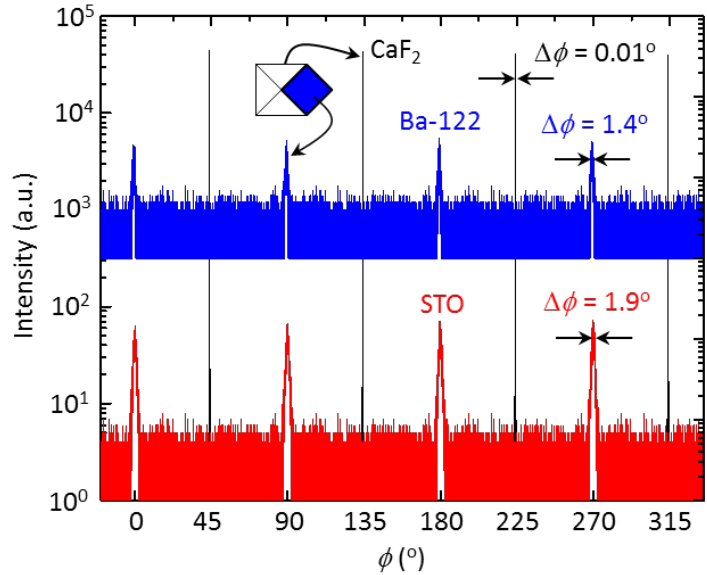

**Fig. S2. In-plane texture and epitaxial quality of Ba-122$_{7\,nm}$/STO$_{14\,nm}$ superlattice system.** Azimuthal ϕ scan and Δϕ of the off-axis (1 1 2) reflections of Ba-122 and STO and the (1 1 1) reflection of $CaF_2$ substrate.

## 2. The relationship between crystallographic structure and tetrahedral coordination

Both tetragonal and orthorhombic structures need to be controlled in the presence of tetragonal-to-orthorhombic transition, and Ba-122 dimension tuning allows the modification of both tetragonal and orthorhombic structures to approach a regular tetrahedron. The reduction of Ba-122 dimension offers a more c-elongated tetragonal structure and less orthorhombic transition due to higher compressive strain and stronger clamping effect (Figure S3a). Those structural changes can drive bond angles $\alpha$, $\beta$, and $\gamma$ equivalent (Figure S3b) and then the regular tetrahedron is successfully achieved.

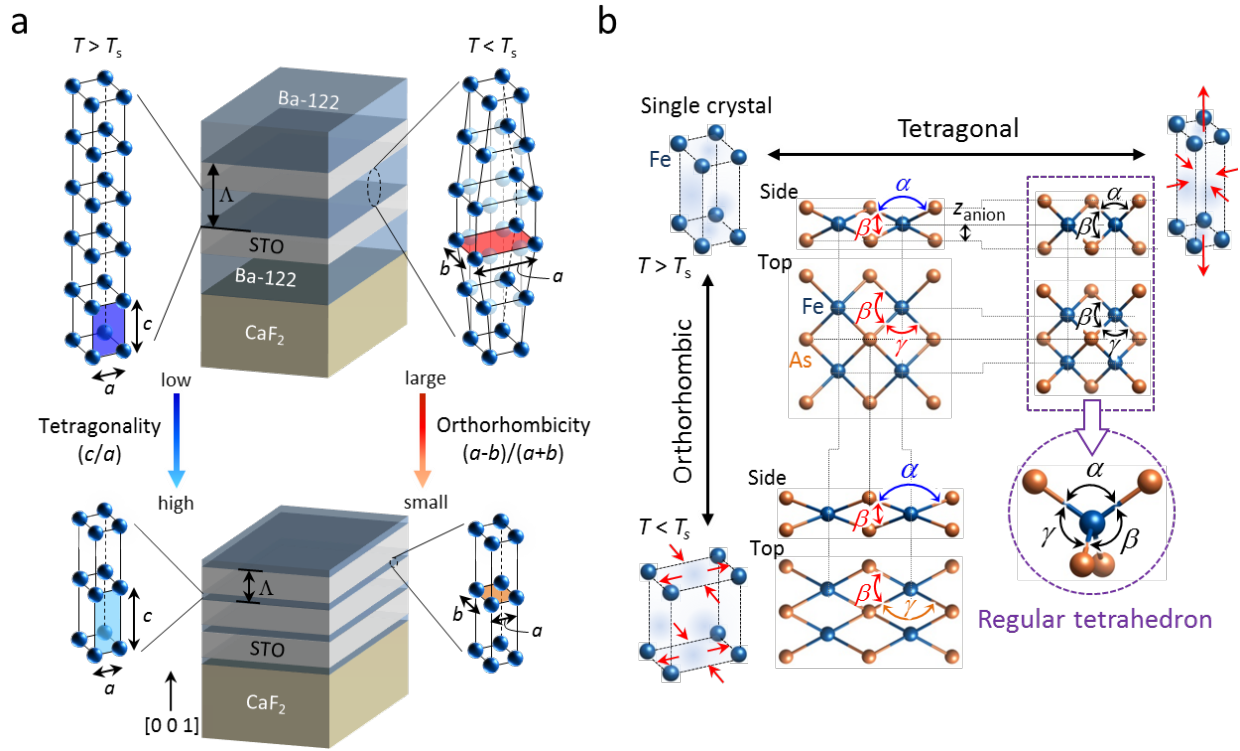

**Fig. S3. The control of crystal structure and tetrahedral geometry by superlattice design.** (a) Ba-122/STO superlattice design for manipulating tetragonal/orthorhombic structures of Fe sublattices. (b) Structural control of Fe sublattices and its influence on tetrahedral geometry.

## 3. Layer thickness and modulation length

High-resolution XRD was performed to study the interference fringe patterns at beamline 6-ID-B of the Advanced Photon Source (APS) at Argonne National Laboratory. Figure S4 clearly shows satellite peaks with the information of each layer thickness and modulation length, in agreement with STEM analysis in Figure S5.

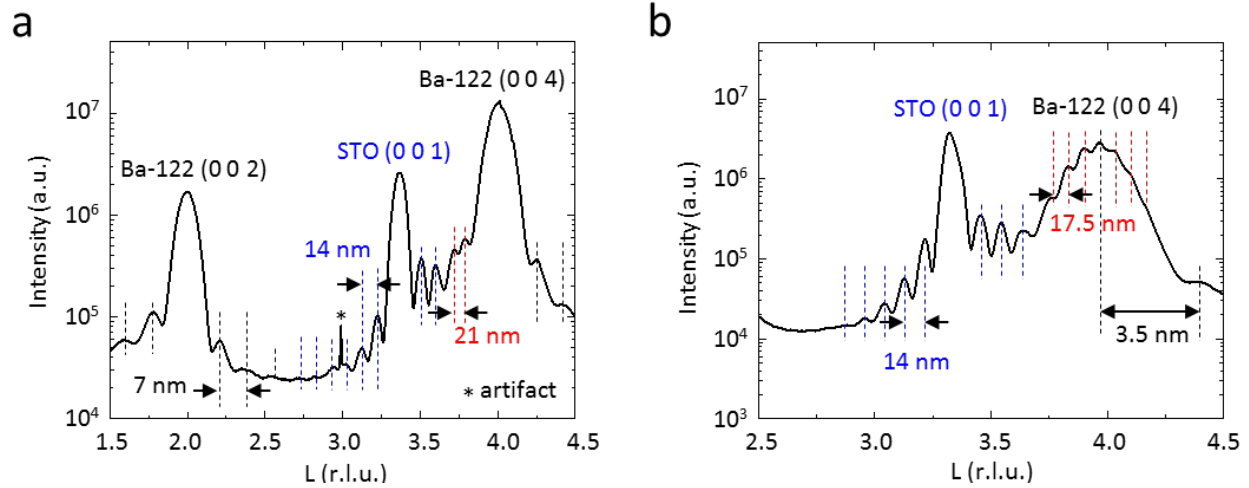

**Fig. S4. Strong satellite peaks in XRD patterns.** X-ray diffraction patterns show the thickness fringe spacing around STO (0 0 1) and Ba-122 (0 0 4) reflections in (a) 7 nm thick and (b) 3.5 nm thick Ba-122 layers in the superlattice. Pronounced thickness fringes indicate the thickness of Ba-122 and STO layers and modulation length ($\Lambda$), corresponding to the STEM imaging.

## 4. Interdiffusion/intermixing at the interface

Since the EDX is useful for heavy elements and Ti-$K_\alpha$ is overlapped with Ba-$L_\alpha$, we also carried out electron energy loss spectroscopy (EELS) scan of O-$K$ and Ti-$L_{2,3}$ edges through the interface. The collection angle and energy dispersion for EELS spectrum imaging are 20 mrad and 0.5 eV/channel, respectively. There is no severe intermixing between Ba-122 and STO in both heavy and light elements.

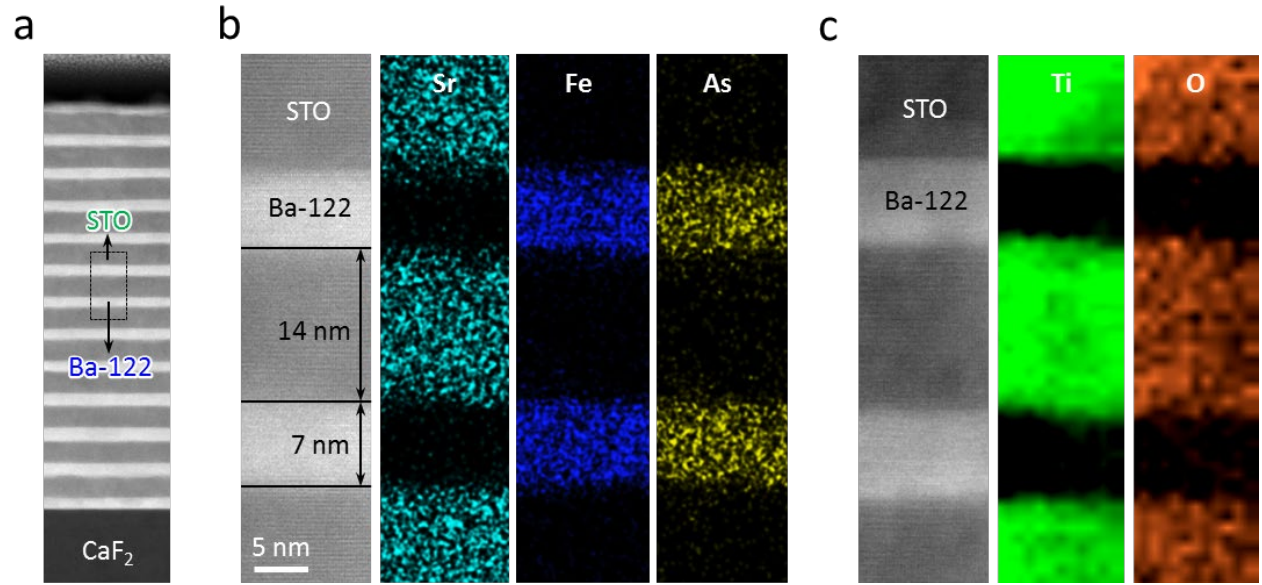

**Fig. S5. Chemical distribution in the Ba-122/STO superlattice.** (a) HAADF image of 12 layers of Ba-$122_{7\ nm}$/STO$_{14\ nm}$ superlattice on CaF$_2$ substrate. (b) EDX mapping for Sr, Fe and As elements (c) EELS scan across the interface for Ti and O elements.

## 5. Tetragonal/orthorhombic structures and FeAs$_4$ tetrahedral geometry

### 5.1. Crystallographic structures

The structural change of Ba-122 is driven by the thermal contraction of the CaF$_2$ substrate, particularly in-plane lattice parameters (Figure S6a). Out-of-plane lattice parameters naturally change by following Poisson's effect (Figure S6b) and tetragonal volumes decrease (Figure S6c).

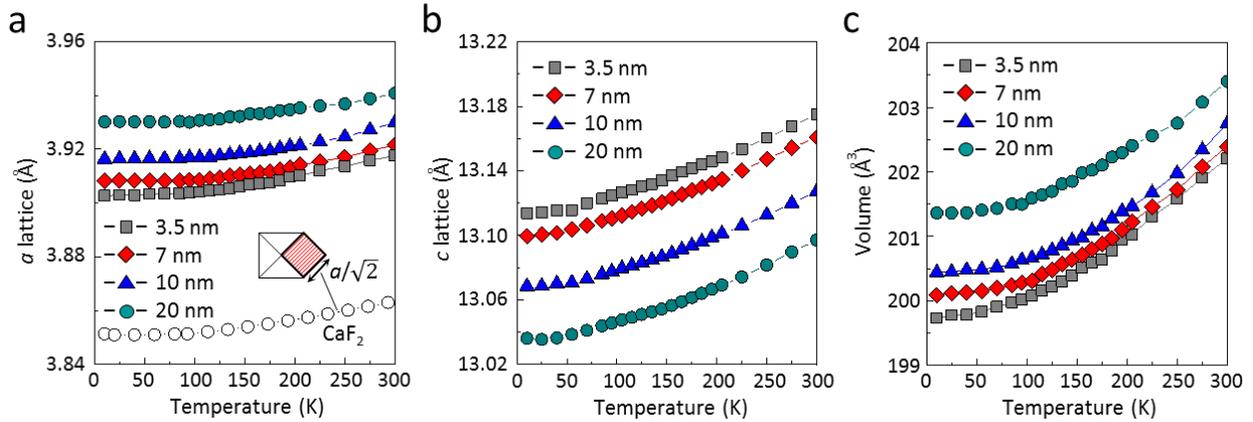

**Fig. S6. Lattice parameters and other structural information by synchrotron XRD.** (a) In-plane lattice parameters in tetragonal notation and (b) out-of-plane lattice parameters of the Ba-122 as a function of temperature. The temperature-dependent in-plane lattice constants are driven by thermal contraction of CaF$_2$ substrates. (c) Calculated tetragonal volumes based on the in-plane and out-of-plane lattice parameters.

## 5.2. Peak broadening

A single sharp peak without the orthorhombic phase is observed above the structural transition temperature ($T_s$). Upon cooling below $T_s$, the (2 2 8) reflection is broadened due to the peak splitting, associated with the orthorhombic transition in bulk Ba-122. We found the full-width half maximum (FWHM) of the (2 2 8) reflection increases due to peak splitting associated with the orthorhombic transition, and the FWHM at room temperature was taken as a reference of tetragonal phase.

Figure S7a shows the (2 2 8) reflections measured at both 300 K and 10 K so that thinner Ba-122 layer (3.5 nm) has less peak broadening rather than thicker Ba-122 layer (20 nm). We measured (2 2 8) reflections as a function of temperature and calculated temperature-dependent ΔFWHM to compare the degree of peak broadening depending on Ba-122 thickness (Figure S7b). It shows that the ΔFWHM gets smaller by reducing Ba-122 thickness.

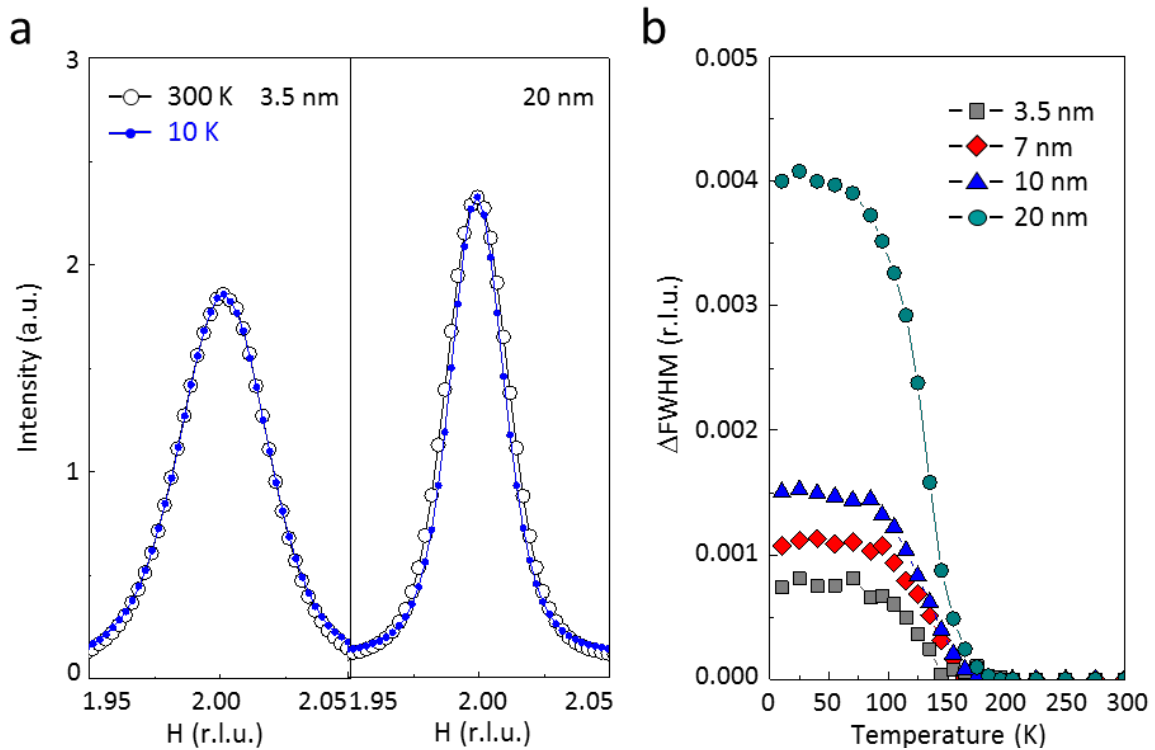

**Fig. S7. Peak broadening associated with the orthorhombic transition.** a) The (2 2 8) reflections of Ba-122$_{3.5 \text{ nm}}$/STO$_{14 \text{ nm}}$ superlattice and Ba-122$_{20 \text{ nm}}$/STO$_{14 \text{ nm}}$ superlattice at 300 K and 10 K. b) Temperature-dependent ΔFWHM depending on Ba-122 thickness.

## 5.3. Tetragonal-to-orthorhombic transition

The (2 2 8) reflection is fitted with two peaks using the FWHM of tetragonal phase (Figure S8a), represented by the purple and dark blue curves indicating the anisotropy of in-plane lattice constants *a* and *b* due to orthorhombic distortion. The in-plane lattice parameters were collected as a function of temperature (Figure S8b), and finally, temperature-dependent orthorhombicity ($\delta$) and tetragonality ($c/a$) can be extracted (Figure S8c and S8d). As the Ba-122 thickness decreases, $\delta$ is suppressed by the clamping effect, and $c/a$ is increased by compressive bi-axial strain.

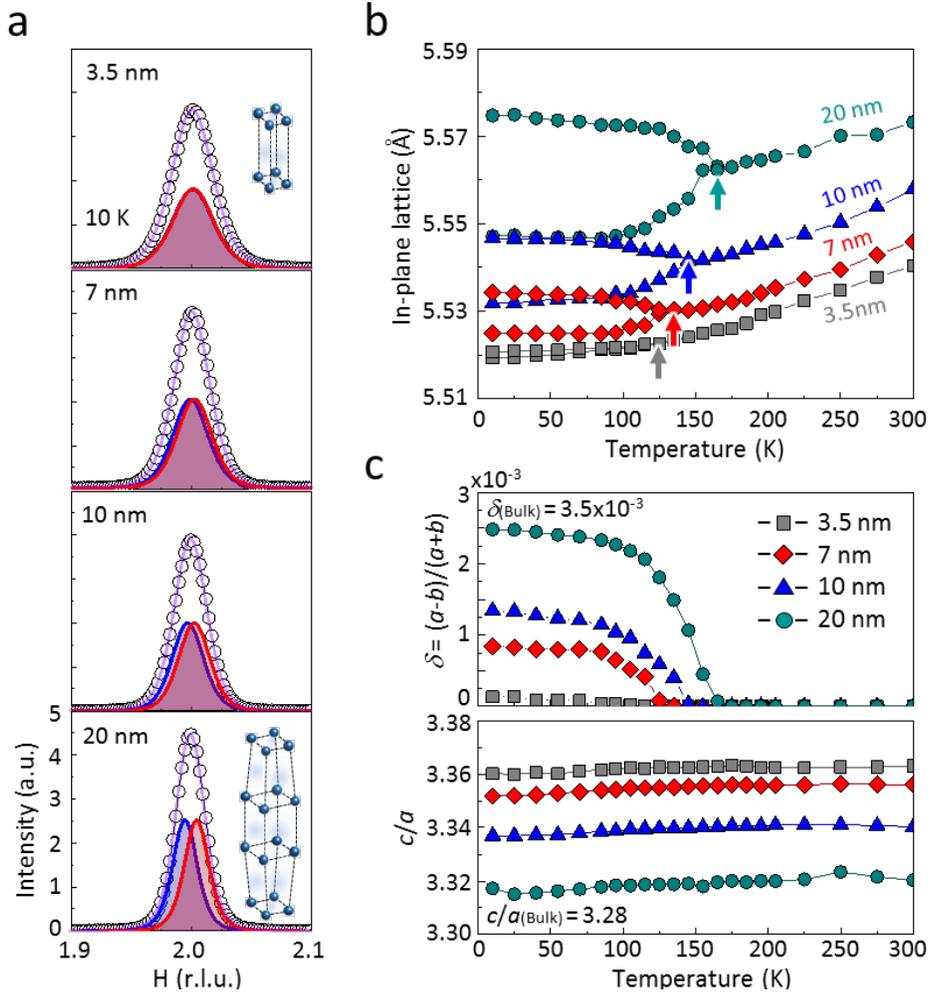

**Fig. S8. Orthorhombic and tetragonal structures of Ba-122 controlled by the Ba-122 dimension.** a) Tetragonal-to-orthorhombic structural change in (2 2 8) reflection by X-ray diffraction at different film thickness. b) In-plane lattice parameters through the tetragonal-to-orthorhombic phase transition. c) Temperature-dependent orthorhombicity ($\delta$) and tetragonality ($c/a$) extracted from (2 2 8) reflections.

## 5.4. Tetrahedral geometry described by anion heights and bond angles

We performed resonant synchrotron X-ray scattering to extract $z$ position of As atoms by comparison with a series of calculated energy scans which are energy dependent at As absorption edges as the resonant scattering terms. With the combination of synchrotron X-ray diffraction and resonant scattering, lattice information and As relative $z$ position was obtained, and then As-Fe-As bond angles were accurately calculated. To keep tetragonal symmetry, only the relative $z$ of As atoms can change in Ba-122 unit cell. Each anion height is closely correlated with each bond angle (see Methods and Figure S9a). From the energy scans for resonant scattering of (0 0 8) reflections, the relative As $z$ position in these samples at 10 K is 0.3552 (Figure S9b). We found there is no difference of the resonant scattering in the temperature range from 10 K to 300 K, which means the relative As $z$ position does not change. Even though in-plane lattice parameters change as a function of temperature, the relative $z$ of As atoms are not sensitive to the change of in-plane strain in the same sample whereas out-of-plane lattice constant is sensitively changed depending on in-plane strain following Poisson's effect. In all samples, there is no difference in resonant scattering spectra despite the different strain states in Figure S9b. However, we need to consider the error range of the relative $z$ value with 0.0012.

With the combination of X-ray diffraction and resonant scattering at 10 K, lattice information and As relative $z$ position was obtained, and then anion heights and As-Fe-As bond angles were calculated with the error range of relative As $z$ value (see Methods and Figure S9c and S9d).

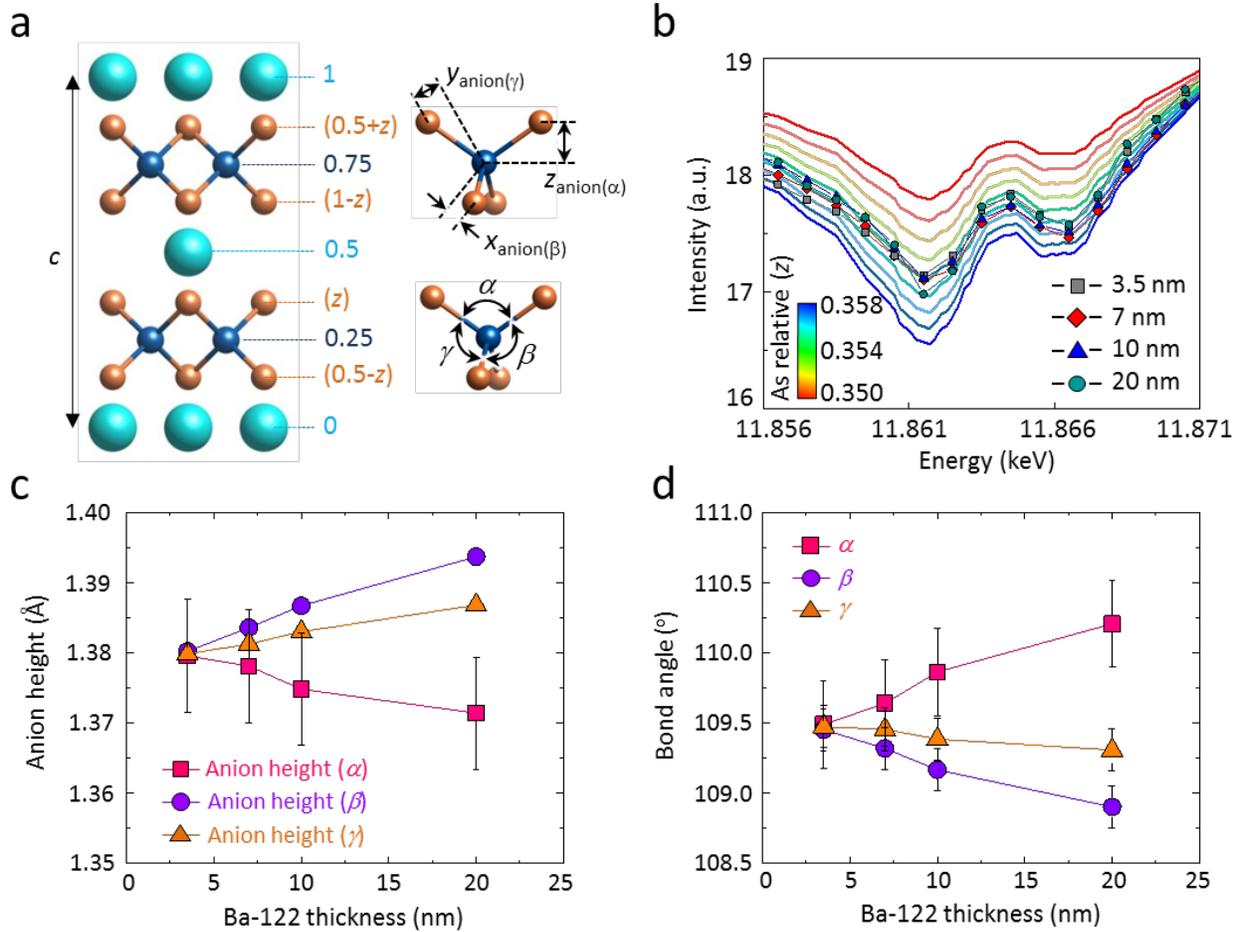

**Fig. S9. Crystal structure and tetrahedron geometry for the calculation of As anion height at 10 K.** (a) Ba-122 tetragonal unit cell and relative $z$ position of Ba, Fe, and As atoms. (b) Energy scan for resonant scattering to determine As relative $z$ position by comparing calculated and measured intensity spectra. (c) Anion heights as a function of the Ba-122 dimension. (d) Thickness-dependent bond angles.

The detailed techniques about bond angle calculations are as follows. The symmetry group of Ba-122 is I4/mmm. Both Ba and Fe are at the special Wyckoff position 2a and 4d respectively with respect to the aforementioned crystal symmetry. As long as the crystal maintains this symmetry with the application of bi-axial strain, the position of the relative positions of Ba and Fe are fixed within the given unit cell. On the other hand, the As position is 4e so that the relative $z$ position (along the $c$ axis) in the unit cell can vary and not break symmetry. For the bulk crystal case under hydrostatic pressure, only As relative $z$ position can be changed from published diffraction data[1-3]. Therefore any intensity difference measured in the (0 0 L) Bragg reflections is only sensitive to changes in As relative $z$ position. Since the proportion of As contribution to these reflections is significant, one can expect to determine the absolute position of the As atom from the As-$K$ edge anomalous X-ray scattering.

Scattering amplitude of the atom can be written in the form

$$f(Q, \omega) = f^0(Q) + f'(\omega) + if''(\omega)$$

where $f'$ and $f''$ are the real and imaginary parts of the dispersion corrections, which are energy dependent at absorption edges as the resonant scattering terms. Basically, using imaginary $f''$ obtained from the fluorescence data, real $f_{As}$ was calculated from the match with (0 0 6) structure factor from the measured intensity of (0 0 6) reflection. Here, the structure factor was used for the given the As relative position so that the $f'$ can vary when you change As position. However, it does not change much because (0 0 6) is not sensitive to the As $z$ position. Therefore, $f_{As}$ was able to be calculated as a constant value and the calculated $f_{As}$ was generated to match the energies for $f''$, with 3 eV offset for the convenience.

$$f_{As}(Q, \omega) = f^0(Q) + f'(\omega) + if''(\omega)$$
$$= \left(F_{(0\,0\,6)} - f_{Ba} \cdot \sum e^{-2\pi i(6 \times (0+0.5))} - f_{Fe} \cdot \sum e^{-2\pi i(6 \times (0.25+0.75))}\right) / \sum e^{-2\pi i(6z)}$$

With the $f_{As}$, another reflection of (0 0 8) was calculated and compared with measured (0 0 8) reflection of the Ba-122 layers.

$$F_{(0\,0\,8)} = f_{As}(Q, \omega) \cdot \sum e^{-2\pi i(8z)} + f_{Ba} \cdot \sum e^{-2\pi i(8 \times (0+0.5))} + f_{Fe} \cdot \sum e^{-2\pi i(8 \times (0.25+0.75))}$$

For all the films with different stain states, the relative As contribution to the (0 0 L) reflections does not change. From the (0 0 8) energy scans shown in Supplementary Section 5.4, the relative As $z$ position in these samples is 0.3552. To attempt to extract $z$ position changes a series of calculated energy scans are plotted which can be employed as an As displacement scale is attached, from which a tabulated angle can be estimated and help refine the angles tabulated from the tetragonal unit cell changes.

With the combination of X-ray diffraction and resonant scattering, lattice information and As relative $z$ position was obtained and anion height and As-Fe-As bond angle can be calculated.

$$z_{anion(\alpha)} = c \times (z - 0.25)$$

$$x_{anion(\beta)} = 0.25 \times a$$

$$y_{anion(\gamma)} = 0.25 \times b$$

$$\alpha = 2 \times \tan^{-1}\left(\frac{\sqrt{\left(\frac{a}{4}\right)^2 + \left(\frac{b}{4}\right)^2}}{z_{anion(\alpha)}}\right) = 2 \times \tan^{-1}\left(\frac{\sqrt{\left(\frac{a}{4}\right)^2 + \left(\frac{b}{4}\right)^2}}{c \times (z - 0.25)}\right)$$

$$\beta = 2 \times \tan^{-1}\left(\frac{\sqrt{\left(\frac{b}{4}\right)^2 + \left(cz - \frac{c}{4}\right)^2}}{x_{anion(\beta)}}\right) = 2 \times \tan^{-1}\left(\frac{\sqrt{\left(\frac{b}{4}\right)^2 + \left(cz - \frac{c}{4}\right)^2}}{0.25 \times a}\right)$$

$$\gamma = 2 \times \tan^{-1}\left(\frac{\sqrt{\left(\frac{a}{4}\right)^2 + \left(cz - \frac{c}{4}\right)^2}}{y_{anion(\gamma)}}\right) = 2 \times \tan^{-1}\left(\frac{\sqrt{\left(\frac{a}{4}\right)^2 + \left(cz - \frac{c}{4}\right)^2}}{0.25 \times b}\right)$$

where $a$, $b$, $c$ are orthorhombic lattice parameters and $z$ is As relative position.

## 5.5. Temperature-dependent anion heights and bond angles

Since every atom position is already determined in tetrahedral geometry, anion heights and bond angles can be calculated as a function of temperature (Figure S10). As the Ba-122 dimension is lower, the tetrahedron approaches optimum values of anion height (0.138 nm) and bond angle (109.5°).

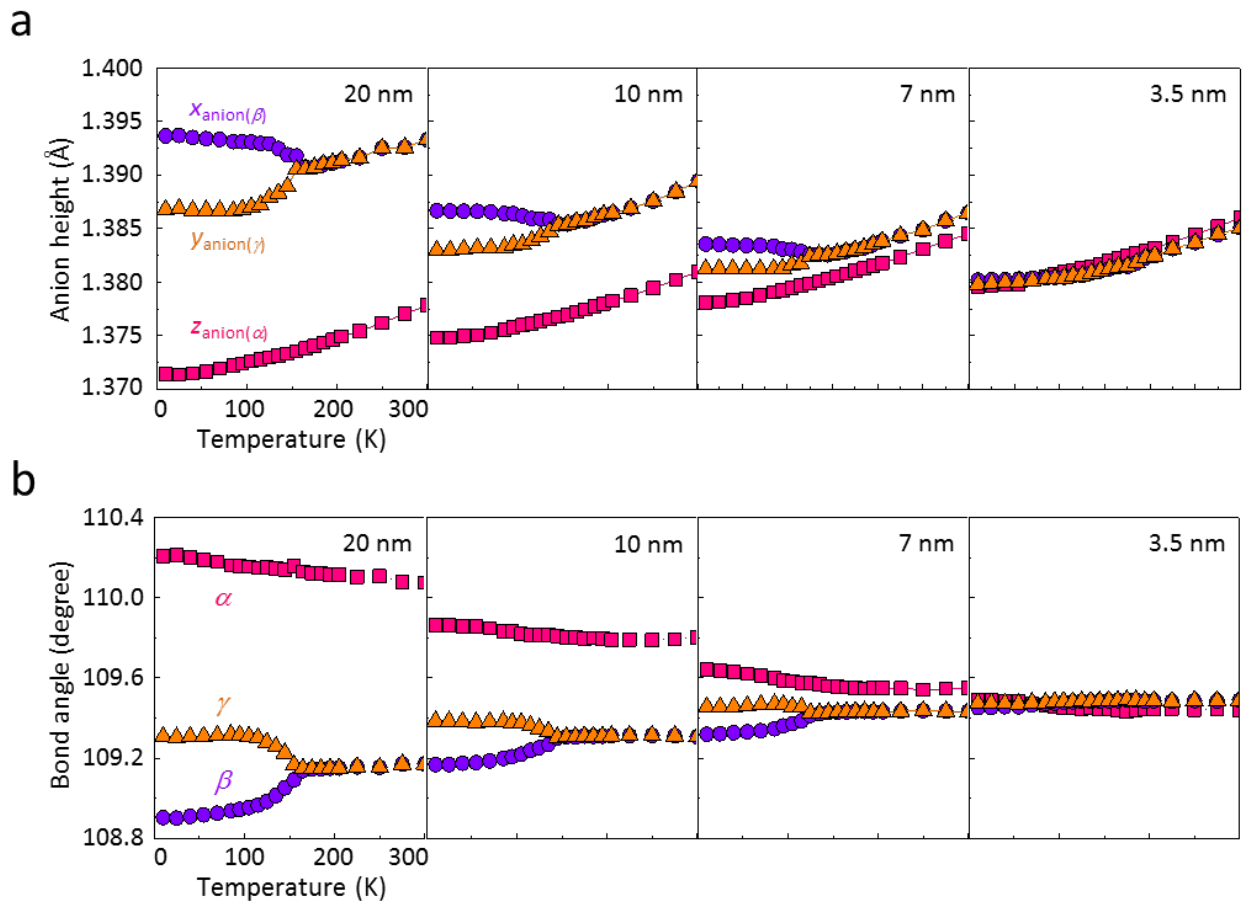

**Fig. S10. Evolution of tetrahedral geometry by Ba-122 dimension control.** Control of tetrahedral geometry as a function of temperature indicated by (a) anion heights and (b) bond angles.

## 6. Effects of Ba-122 and STO dimensions on transport measurements

### 6.1. Ba-122 dimensions

We performed the first derivative of resistivity to precisely check $T_s$ and $T_n$ and to see the relationship between tetrahedral geometry and electromagnetic properties. $T_s$ and $T_n$ are obtained from the onset and maximum points of the first derivative resistivity, respectively (Figure S11). Both $T_s$ and $T_n$ are decreased by the proximity to the regular tetrahedron, indicating that structural control can affect electrical transport ($T_s$) and magnetic ordering ($T_n$). We have tried to detect AFM ordering directly around magnetic Bragg peak (0.5, 0.5, 3) which is supposed to be strong in Ba-122 single crystal using the SPINS cold-neutron triple-axis spectrometer at the NIST Center for Neutron Research. No sign of magnetic peak and abnormal temperature-dependence has been observed in the superlattice thin films because the sample volume may not be as big as a single crystal to detect antiferromagnetic ordering and the magnetism may be already reduced by dimension control.

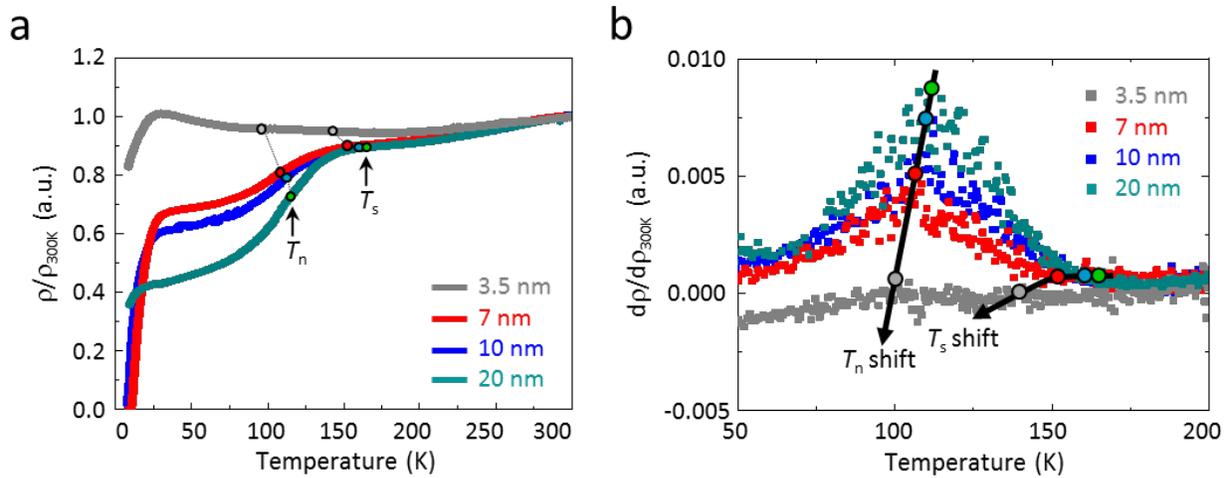

**Fig. S11. Deterministic $T_s$ and $T_n$ from resistivity.** (a) Temperature-dependent normalized resistivity with different Ba-122 dimension. (b) The first derivative of the resistivity of the films to clearly see the shifts of $T_s$ and $T_n$ to lower temperature as the film thickness decreases.

## 6.2. STO dimensions

STO plays an important role to lock the orthorhombic transition of Ba-122 in-plane lattices at both top and bottom interfaces. We have changed the thickness of the STO layer and measured the resistivity to see the shift of $T_s$ and $T_n$ (Figure S12a and S12b). When the STO is too thin, the STO layer is not stiff enough to prevent the orthorhombic distortion of Ba-122. As the STO thickness increases, both $T_s$ and $T_n$ are decreased due to stronger clamping effect and finally saturated above 14 nm thickness. We have fixed the STO layer as thick as 14 nm to study the effect of the Ba-122 dimension.

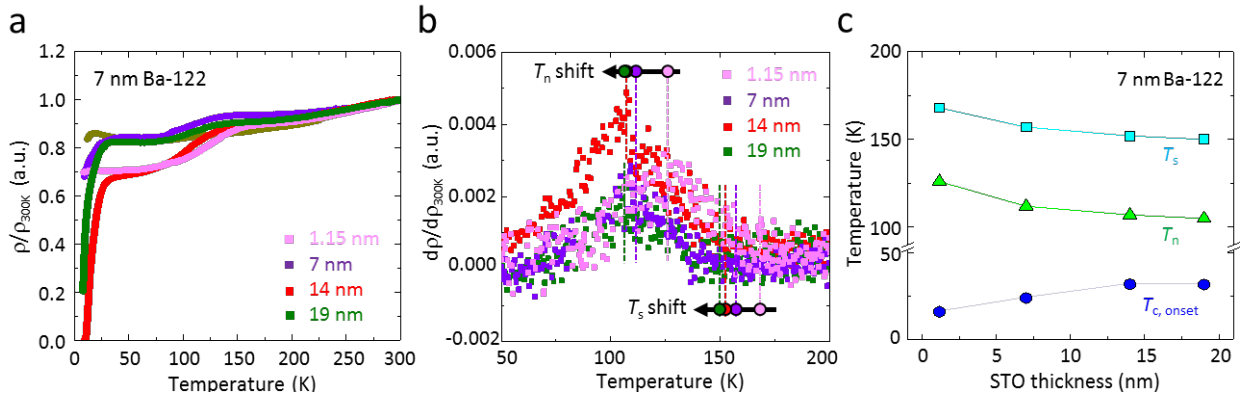

**Fig. S12. Optimization of STO thickness for the study of the clamping effect.** (a) Temperature-dependent normalized resistivity and (b) the first derivative of the resistivity of the 7 nm thick Ba-122 with different STO dimension. (c) $T_s$ and $T_n$ shifts as a function of STO thickness. $T_s$ and $T_n$ are decreased by increasing STO thickness because thicker STO is more effective to clamp the lattice of Ba-122 and to prevent orthorhombic transition. Above 14 nm thickness, no more suppression of $T_s$ and $T_n$ is observed, and $T_{c,onset}$ is also saturated.

# 7. Electromagnetic properties

## 7.1. Sheet resistance and resistivity

We compared sheet resistance and resistivity with different Ba-122 thickness in thin films (Figure S13a and S13b) and superlattices (Figure S13d and S13e) grown on the $CaF_2$ substrates. We only consider total Ba-122 thickness because the resistivity of the STO layer is relatively higher than that of Ba-122. The thickness dependence of sheet resistance and resistivity shows that both thin films and superlattices have the same trend indicating that the most current flows through Ba-122 layers (Figure S13c and S13f). However, only the resistivity of superlattices is higher at 3.5 nm thick Ba-122 layer (blue dotted circle) because of the possible scattering at reduced dimensions. Since the coherence length along the c-axis of the Ba-122[4-7] has been reported in the range from 0.64 nm to 2.45 nm, which is smaller than our minimum thickness, there could be extrinsic scattering in the Ba-122$_{3.5\ nm}$/STO$_{14\ nm}$ superlattice due to the interfacial roughness such as chemical intermixing or atomic vacancies.

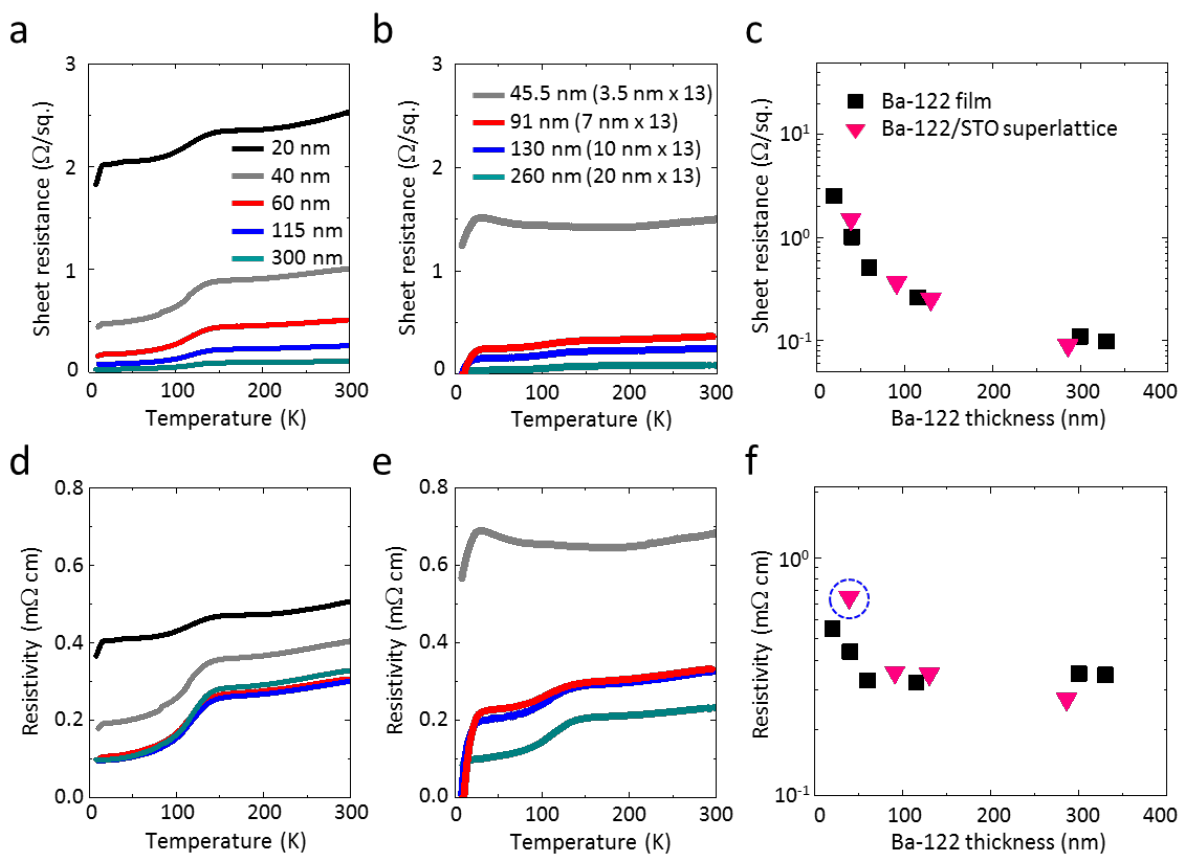

**Fig. S13. Electrical transport properties of Ba-122/STO superlattices.** The sheet resistance of (a) Ba-122 thin films on the $CaF_2$ and (b) Ba-122/STO$_{14\ nm}$ superlattices films on the $CaF_2$. (c) The sheet resistance of the Ba-122 films and Ba-122/STO superlattices as a function of Ba-122 thickness. Ba-122 thickness-dependent resistivity of (d) Ba-122 thin films and (e) Ba-122/STO superlattices. (f) The resistivity of Ba-122 films and Ba-122/STO superlattices as a function of Ba-122 thickness.

## 7.2. Resistive transition with the magnetic field

Resistive broadening is observed by increasing the magnetic field, but the broadening is not big enough to show ohmic dissipation in Ba-122/STO superlattices due to strong pinning behavior even in parent Ba-122 (Figure S14).

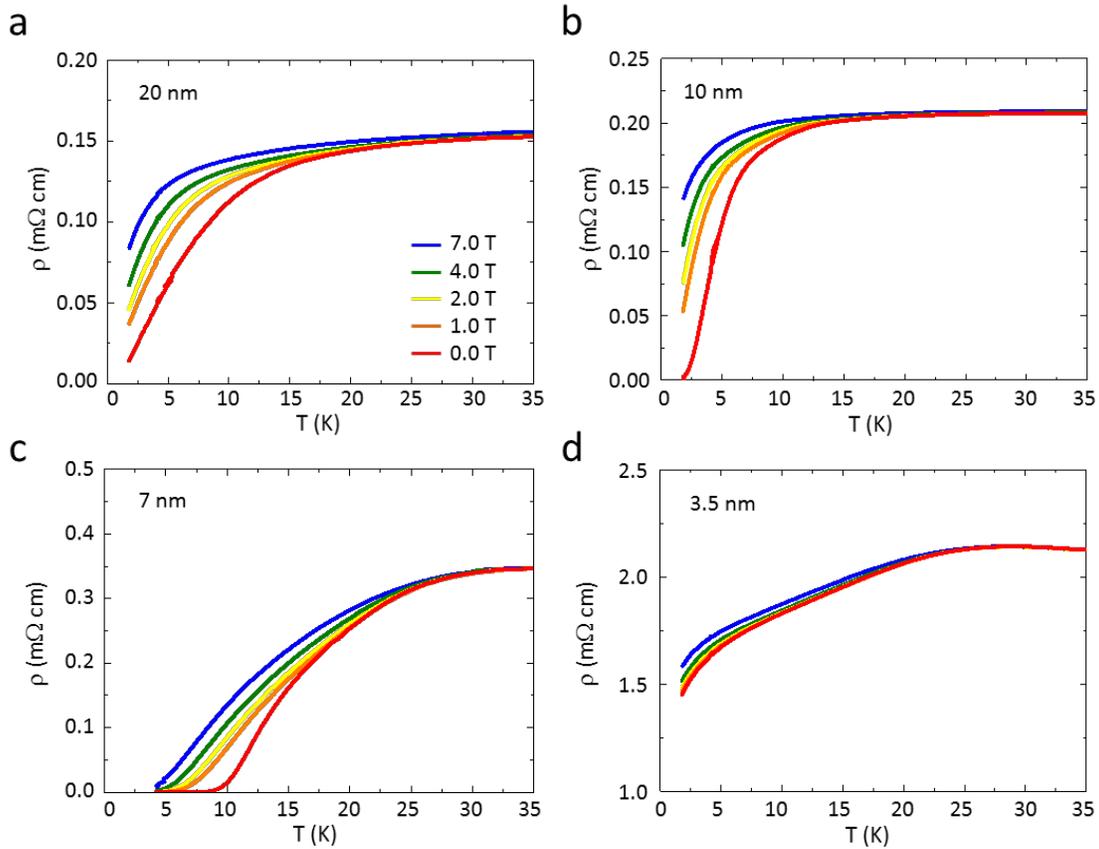

**Fig. S14. Superconducting characteristics of Ba-122/STO$_{14\,nm}$ superlattices.** The superconducting resistive transition of (a) 20 nm, (b) 10 nm, (c) 7 nm and (d) 3.5 nm thick Ba-122 with different perpendicular magnetic field strengths.

## 7.3. V-I characteristics and magnetic moment

The existence of superconductivity was also confirmed by voltage-current characteristics and magnetization measurements. As temperature decreases, ohmic behavior disappears and random noise has been measured at a low current region to show the superconducting state (Figure S15a). The magnetization $T_c$ determined from a superconducting quantum interference device (SQUID) measurement indicates the onset of bulk superconductivity, whereas superconducting $T_c$ measured by transport measurements represents the formation of percolation paths (Figure S15b).

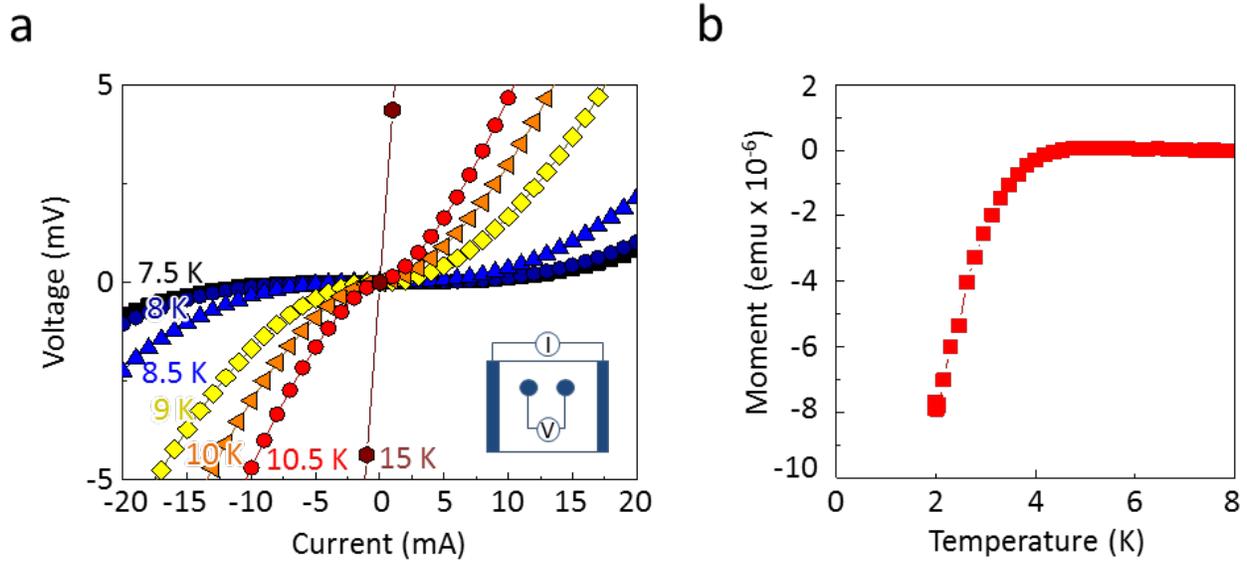

**Fig. S15. Superconducting properties of Ba-122$_{7\,nm}$/STO$_{14\,nm}$ superlattice.** (a) Voltage-current characteristics at different temperatures. (b) Magnetic moment as a function of temperature. A field of 100 Oe was applied perpendicular to the plane of the films after cooling to 2 K.